# Situating the social issues of image generation models in the model life cycle: a sociotechnical approach[*]

Amelia Katirai[1], Noa Garcia[2], Kazuki Ide[1,3], Yuta Nakashima[2], Atsuo Kishimoto[1,2]

[1] Research Center on Ethical, Legal, and Social Issues, Osaka University
[2] Institute for Datability Science, Osaka University
[3] Center for Infectious Disease Education and Research, Osaka University

**Abstract**
The race to develop image generation models is intensifying, with a rapid increase in the number of text-to-image models available. This is coupled with growing public awareness of these technologies. Though other generative AI models—notably, large language models—have received recent critical attention for the social and other non-technical issues they raise, there has been relatively little comparable examination of image generation models. This paper reports on a novel, comprehensive categorization of the social issues associated with image generation models. At the intersection of machine learning and the social sciences, we report the results of a survey of the literature, identifying seven issue clusters arising from image generation models: data issues, intellectual property, bias, privacy, and the impacts on the informational, cultural, and natural environments. We situate these social issues in the model life cycle, to aid in considering where potential issues arise, and mitigation may be needed. We then compare these issue clusters with what has been reported for large language models. Ultimately, we argue that the risks posed by image generation models are comparable in severity to the risks posed by large language models, and that the social impact of image generation models must be urgently considered.

## 1. Introduction

Rapid progress in image generation models has caused major disruption, both to the field of computer science, and to society at large.  Within the span of just a few months in 2022, multiple text-to-image models, including DALL-E 2 (1) (April 2022), Dall-E Mini (2), Midjourney (July 2022), and Stable Diffusion(3) (August 2022) were released, with the ability to produce high-quality photo-realistic and artistic images from a text prompt. These developments were rapidly followed by the release of text-to-video models including CogVideo (4), Make-A-Video (5), or Imagen Video (6). The high quality of the visual output generated, easy-to-use interfaces, and powerful commercial incentives have pushed these models out of research labs and into the awareness of the public. However, the speed of these advancements has created a vacuum, as there remains relatively little understanding or awareness of the social implications of these shifts. Where there is broader awareness of social issues, the attention is often unbalanced, with a tendency, for example, to prioritize issues related to intellectual property, while other issues remain insufficiently understood.

---

[*] This preprint has not undergone peer review or any post-submission improvements or corrections. The Version of Record of this article is published in *AI and Ethics*, and is available online at https://doi.org/10.1007/s43681-024-00517-3.

Although interest in the use of computers to generate art was present throughout the 1900s, traced back as far as to the 1950s, the technology ultimately reached a breakthrough in 2015 through Google's development of DeepDream, a neural network trained on over 10,000 paintings (7). Earlier image generators were based on a type of machine learning referred to as generative adversarial models (8), but as will be described below, current text-to-image models leverage CLIP (9) embeddings to jointly capture language and visual semantics, drawing on CLIP's general-purpose image-text model (7). Both CLIP and the image generation models based on it owe their high performance to a training process based on massive collections of data crawled and scraped from the Internet, which allows the models to classify images without necessarily requiring human labeling (7). At the same time, the use of such uncurated data and its effects on the final model's output raises both immediate and longer-term social and ethical concerns, including risks of bias, threats to intellectual property and privacy, and broader impacts on the social environment.

To date there has been relatively little consideration of the social issues raised by image generation models, with the exception of some recent groundbreaking work in the field (10,11). In this paper, we provide a comprehensive categorization of the risks posed by the mass-scale adoption of these models. The concerns around image generation models share similarities with other types of generative models such as LLMs; however, the particularities and implications of the visual modality deserve to be paid special attention. Thus, as with the ongoing scholarly debate about large language models (LLMs)(12), our aim is to raise awareness about the potential negative societal impact of image generation models.

Joyce et al. (13) have noted a need for increased consideration of the sociotechnical nature of machine learning models, attending to both the technical aspects of the models, and their social aspects, drivers, and implications. Conceptualizing this as the intersection of two axes—a technical axis and a social axis—we find that the relatively limited work to date on image generation models has been situated along either one of these axes. We report here on the results of a novel interdisciplinary collaboration at the *intersection* of these two axes, between experts in machine learning and those in the social sciences. Drawing on a review of the recent literature, as well as this interdisciplinary expertise, we first detail the components in the life cycle of image generation models, from how the training data is acquired to the long-term use of the generated output. Then, we offer a novel taxonomy for the potential social issues of image generation models. By intersecting the technical and social axes, we identify where in the model life cycle of the technology each issue arises or is reinforced. Through this, we ground the longer-term—and at times more abstract—issues of these models within the models themselves, while also providing a guide both to the model life cycle, and to its associated issues, for researchers working along either of these axes. It is noteworthy that, although we use "social" here as an overarching category, we refer broadly here to non-technical matters, including issues which have been framed elsewhere as ethical ones, as well as those from the humanities.

To this end, along the technical axis, we identify four main phases in the model life cycle: data acquisition, model design, deployment, and the use of generated images; along the social axis, we identify seven issue clusters: data issues, intellectual property, bias, privacy, and the impacts on the informational, cultural, and natural environments, respectively. Given the strong recent attention to LLMs, we use LLMs as a point of comparison, comparing the risks identified for image generation models with those reported for LLMs. We find that, despite the relative lack of attention to the risks of image generation models, their risks are comparable in severity. Issues of intellectual property, bias, and the impact on the information and cultural environments are especially of concern, given the unique features of images as a medium. As a result of this sociotechnical examination, we argue that attention to these profound risks is urgently needed by researchers along both axes, and at their intersection. We close with preliminary recommendations to minimize the negative impact of text-to-image models in society. Throughout, the paper contributes to the early body of literature on this theme by providing a summary of the key issues surrounding text-to-image models, while at the same time breaking ground by situating these social issues within the model-life cycle of text-to-image models.

## 2. The model life cycle

Text-to-image models generate images based on an input text. Here, we examine the life cycle of such models. The first models creating images from text were proposed back in 2016 (14) by encoding an input sentence through a long short-term memory network (LSTM) that conditioned an image decoder based on a convolutional neural network (CNN). Datasets used for training such networks were rather small, limiting the model's capabilities and, as a result, producing notably blurry images. Later, with DALL-E (15), a breakthrough in image quality was achieved by leveraging CLIP (9) embeddings. As a method for encoding images and texts, CLIP learns the semantic correspondence between the visual appearance of objects in images and the words in text by comparing image and text data samples in a collection of about 400 million image-text pairs from the Internet. Current text-to-image models build on top of the semantic knowledge of these pre-trained CLIP embeddings to replicate the appearance and composition of the elements mentioned in the input text and generate high-quality images through copious amounts of training data.

Regardless of the obtained image quality, we identify four phases involved in all text-to-image models' life cycle, as shown in Figure 1, which are 1) data acquisition, 2) model design, 3) deployment, and 4) use of generated images; each phase is further divided into the sub-processes defined below.

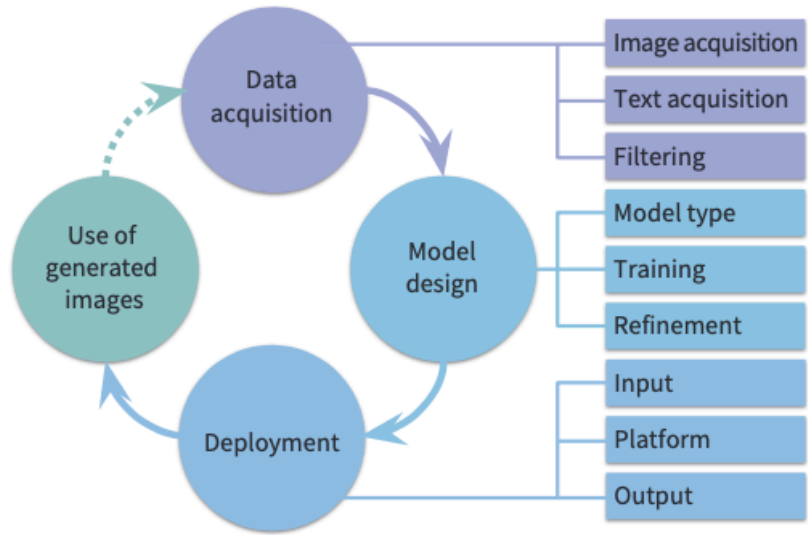

*Figure 1* The model life cycle

*Table 1* Datasets used in text-to-image models

| Dataset | Year | Image acquisition | | Text acquisition | | | Filtering (automatic unless otherwise stated) | Data consent/License |
|---|---|---|---|---|---|---|---|---|
| | | **Number** | **Method** | **Number** | **Method** | **Language** | | |
| SBU Captions (16) | 2011 | 1M | Flickr | 1M | Flickr | English | Text quality | No |
| Flickr30K (17) | 2014 | 0.032M | Flickr | 0.159M | Crowdsourcing | English | Text quality | |
| MSCOCO (18) | 2015 | 0.33M | Flickr | 1.5M | Crowdsourcing | English | Image content | No |
| Visual Genome (19) | 2017 | 0.11M | | | | English | | |
| GCC3M (20) | 2018 | 3.3M | Web crawler | 3.3M | Web crawler | English | Image and text format Offensive content Text-image matching | No |
| Nocaps (21) | 2019 | 0.015M | Flickr | 0.166M | Crowdsourcing | English | None | CC BY 2.0 |
| TextCaps (22) | 2020 | 0.028M | Flickr | 0.142M | Crowdsourcing | English | Image content Manual text quality | CC BY 2.0 |
| GCC12M (23) | 2021 | 12.4M | Web crawler | 12.4M | Web crawler | English | Image and text format | No |

| | | | | | | Offensive content Text-image matching | |
| --- | --- | --- | --- | --- | --- | --- | --- |
| WIT (24) | 2021 | 11.5M | Wikipedia | 37.5M | Wikipedia | 108 lang. | Image format Text quality Offensive content | Research license |
| RedCaps (25) | 2021 | 12M | Reddit | 12M | Reddit | English | Offensive content | Public form to request removal |
| LAION-400M (26) | 2021 | 400M | Web crawler | 400M | Web crawler | English | Image and text format Text-image matching | No |
| LAION-5B (27) | 2022 | 5850M | Web crawler | 5850M | Web crawler | +100 lang. | Image and text format Text-image matching | No |

## 2.1 Data acquisition

As with any machine learning application, text-to-image models need data samples to learn the internal representations of the model and conduct the task of interest. The data acquisition phase consists of gathering and preprocessing those data samples. Text-to-image models require data to be in the format of image-text pairs—for each image there should be an accompanying text with its description. Data samples are used for training, fine-tuning, or evaluating text-to-image models. An overview of data sets used in this phase of the model life cycle is provided in Table 1. Within the data acquisition phase, we identify three sub-phases:

1) **Image acquisition**: The sub-phase for obtaining images. Early image-text datasets were mostly domain-specific and relatively small, with images carefully picked from the web or captured by the dataset authors (e.g., Oxford-102 Flower(28), Caltech-UCSD Bird (29)). In medium-scale and general-domain collections, such as MS-COCO (30), images were obtained by keyword search and filtered during the annotation process. Lately, text-to-image models are trained on web-scale collections of images crawled from millions of websites and with little-to-no human intervention, such as LAION-5B (27).
2) **Text acquisition**: The sub-phase for obtaining text describing the images. In small- and medium-scale image datasets, text can be written by human annotators (31). Human annotations are not feasible in web-scale collections, so the text is commonly obtained together with the images from the alt-text HTML attribute in the original websites (20).
3) **Filtering**: The sub-phase for discarding data samples that are not appropriate for the task. Filtering can be based on multiple criteria, including data domain (e.g. only keeping images from flowers (28)), image quality (e.g., removing images with dimensions smaller than 400 pixels (20)), image or text content (e.g., deleting samples labelled as "not safe for work" or NSFW (25)), license (e.g., only images with research permissive license (24) etc.) In small- and medium-scale datasets, there can be a human verification step to check the quality of the text (22). In web scale datasets, however, automatic filters are applied (20,22).2

## 2.2 Model design

Different architectures, training strategies and parameters can be used to develop a text-to-image generation model. We refer to the process of choosing these components as the model design phase.

1) **Determination of model type**: The choice of the core technology used in the text-to-image model. Early models (e.g. (31)) were based on generative adversarial networks (GANs) and built on top of recurrent neural networks (RNN)-based text encoders and convolutional neural networks (CNN)-based image decoders (8). In most of the newest models, text is encoded with a CLIP (9) pre-trained Transformers (32) CLIP embeddings, and images are decoded either with autoregressive (33,34) and diffusion (35–37) techniques.
2) **Training process**: The process for automatically adjusting (or learning) a model's parameters according to a target task. There are several factors involved in the training process:
    a. **Optimization objective**: Text-to-image models are commonly optimized to increase image quality by setting the target task to make, given an input text, the generated image be as similar as possible to the image originally paired with said text (hence, the necessity of paired image-text samples in the data acquisition process). The target task used during training is a design choice by itself, and by optimizing models based solely on image quality, other variables such as efficiency, cost, or fairness tend to be overlooked.
    b. **Training duration**: Models learn to adjust their internal parameters by iterating over training data and fitting the target task for each of the training samples. In large models, the training process with thousands of millions of text-image pairs can take several months with large computational costs. Information about this processis rarely available , and only DALL-E mini and Stable Diffusion have disclosed this information. According to the model card[†], Stable Diffusion v1.1 was trained for 150,000 GPU hours and emitted 11,250 kg CO2 eq.
    c. **Hyperparameter search**: In addition, models usually have multiple hyperparameters that need to be set manually and are not learnable through training, e.g., learning rate, batch size, number of training steps, etc. It is common to train multiple models with different hyperparameters to find the optimum configuration, increasing computation time and cost. This hyperparameter search is usually not reported in the computational cost of the models.
3) **Refinement**: Although not yet reported in text-to-images models, LLMs have been improved by the use of reinforcement learning (RL) techniques with annotations on whether the generated data is good (38). This process could be potentially applied to text-to-image models to improve the performance.

---

[†] https://github.com/CompVis/stable-diffusion/blob/main/Stable_Diffusion_v1_Model_Card.md

## 2.3 Deployment

Once a model is trained, the next phase is to share it and make it available to users. We refer to this phase as the deployment phase.

1) **Input**: The input to the model consists of a text, also known as a prompt. The input text is fed into the model to generate the output.
2) **Platform**: Where the model is hosted: private server (e.g., DALL-E 2), third-party server (e.g., DALL-E mini), or the user's own machine (e.g., CogView, Stable Diffusion).
3) **Output**: The output consists of the generated images. For an identical input text, different models can generate different images according to each model's style. The use of such output is listed as a separate entry of the model life cycle because of their importance.

## 2.4 Use of generated images

The final phase is the use of the generated images, which may be utilized by users in a range of settings. However, as the generated images can be published on the Internet, they can also potentially be collected in the data acquisition phase to train future models.‡ We see two key avenues through which this can occur.

1) **By users**: As the generated images can be published back on the Internet, they can potentially be collected in the data acquisition part for training future models.
2) **By the model developer**: Improvement of the model by annotating the generations. Large language models have been reported to use reinforcement learning for this task and could be applied to text-to-image models in the future (38).

Thus, despite the diversity in the range of text-to-image models available, they share commonalities in the steps across tehir life cycle. A summary of the major text-to-image models is provided in Table 2.

*Table 2* General text-to-image models. There are two main families of models: autoregressive (AR) or diffusion (Diff.). Access: X neither the code nor the model has been released; ✓ the model can be freely accessed; 🔑 the model is only accessible under registration.

| Model | Date | Institution | Type | Params(B) | Images (M) | Access |
|---|---|---|---|---|---|---|
| DALL-E | 2021/02 | OpenAI | AR | 12.0 | 250 | X |
| CogView | 2021/06 | Tsinghua Uni. | AR | 4.0 | 30 | ✓ |
| GLIDE | 2021/12 | OpenAI | Diff. | 5.0 | 250 | X |
| GLIDE (filtered) | 2021/12 | OpenAI | Diff. | 0.3 | 137 | ✓ |
| CogView 2 | 2022/03 | Tsinghua Uni. | AR | 24.0 | 30 | ✓ |
| DALL-E mini | 2022/04 | Craiyon | AR | 0.4 | 15 | ✓ |
| DALL-E 2 | 2022/05 | OpenAI | Diff. | 6.5 | 650 | 🔑 |
| Imagen | 2022/06 | Google | Diff. | 3.0 | 860 | X |
| Parti | 2022/06 | Google | AR | 20.0 | 9 | X |
| Midjourney | 2022/07 | Midjourney | - | - | - | 🔑 |
| Stable Diffusion (v1.1) | 2022/08 | StabilityAI | Diff. | 1.5 | 2,000 | ✓ |
| Firefly | 2023/03 | Adobe | - | - | - | 🔑 |

---

‡ This is why we have connected "Generated images" and "Data acquisition" with a dotted arrow in Figure 1.

# 3. Social issues in image generation systems

Following on this map of the model life cycle, we report on a review of the literature on the social issues arising from image generation systems. Expanding on existing approaches to image generation systems by Zhang et al. (39) and Ko et al. (40) and with acknowledgement of recent work on image generation systems (10,11) and reference to recent work on LLMs (12,41), we present seven issue clusters: data issues, intellectual property, bias, privacy, and the impacts on the informational, cultural, and natural environments.

The review undertaken here was based on a flexible review method. A flexible approach was prioritized over a structured or systematic approach to accommodate the rapid developments in the field. The review drew primarily on hand-searches and used a snowballing approach to expand from core papers in the field to other related works. This was supplemented with independent searches of the academic literature and grey literature, and reference sharing within the interdisciplinary team to identify relevant works from various fields. Throughout, we prioritized most-recent works. The review here is non-exhaustive but intended to convey key themes from the literature. Thus, the literature search was brought to a close once saturation had been perceived to be reached given the scope of the paper.

These issue clusters were extracted from the literature using an approach informed by thematic analysis (42). Key areas of discussion in the literature were first identified and noted. These were then grouped into broader thematic categories which we refer to here as issue clusters. The issue clusters were reviewed and adjusted by the interdisciplinary team in light of their expertise. It is noteworthy that there are novel initiatives across multiple areas to bring progress on addressing or resolving some of the issues identified. However, issues were taken as the focus of this review, and such initiatives were treated as outside the scope of this exploratory analysis. Each issue cluster and noteworthy points raised in relation to it will be taken up in turn in the sub-sections that follow.

## 3.1 Data issues

The first issue cluster identified is data-related issues. Image generation models are reliant on massive amounts of "unconstrained data"—datasets gathered through means such as web scraping. These methods, though cost-efficient, involve "regarding people as `raw material' without the right to privacy, or the agency to consent or opt-out" (43). Gebru et al. and others (43,44) highlight the necessity of appropriate documentation for datasets used in machine learning, to ensure that harms can be avoided or remedied. Yet, the size of the datasets used in the development of generative models poses an inherent obstacle to such mitigation, particularly as they lack metadata which could be used as ground truth for their evaluation (43). This is compounded as the creators of the models may be disincentivized from making public the details of their training data, due to the major investments involved in their creation (45). Therefore, the identification of emergent issues related to training data often relies on the analysis of output. For example, Schramowski et al. (46) analyzed output generated using Stable Diffusion, finding "a significant amount of inappropriate content generated which, viewed directly, might be offensive, ignominious, insulting, threatening or might otherwise cause anxiety," highlighting problematic training data as a key issue. This points to inherent issues in this approach, where the size of the data set prohibits documentation. This has led some scholars to critique the way in which larger datasets are "mythologized as beneficial" (43) for training models, proposing rather that smaller datasets and thorough documentation of their contents may help to offset some of the issues arising from them, as will be discussed in multiple sections below.

It is also significant that a lack of documentation can hamper efforts to filter out problematic data from the training sets. Yet, filtering itself can become a site for the reproduction of multiple types of social inequality. For example, Monea (47) argues that content moderation can lead to the creation of a "digital closet," through which LGBTQ+ people are "so pornographied" through content moderation that their sexual expression is restricted. Ungless et al. (48) identify a tendency to "consistently" and problematically represent "certain non-cisgender identities... as less human, more stereotyped and more sexualized," including through the generation of pornographic images. Yet, their study also found that when NSFW images were blocked, this was "likely to contribute to the erasure of non-cis identities," rather than to "prevent the generation of harmful output." In a similar vein, based on their attempts at bias mitigation, Nichol et al. (49) "hypothesize" that the training data used for DALL-E 2 "may be biased toward presenting women in more sexualized contexts," but found that filtering data to remove this content led to underrepresentation of women in output in response to neutral prompts. Further, the content moderation process can itself reproduce social inequality as reported by Gray and Suri (50) for AI broadly, who highlight the precarious employment conditions in which content moderators engage in often traumatic work. In early 2023, Perrigo (51) drew attention to these issues in relation to generative models particularly, raising alarm over the poor working conditions of Kenyan content moderators for OpenAI's ChatGPT—issues which are also reflected in the data annotation process (43).

## 3.2 Intellectual property

Data issues are also closely linked with issues of intellectual property, which can arise both through the training data used for image generation models, and through the output generated by the systems, which may (intentionally or unintentionally) have similarities to copyrighted work (52). It is notable here that understandings of "fair use" differ across national contexts: in the United States, for example, published works may be used for training models, while in the EU, artists are able to disallow the use of their work if it will ultimately be used commercially (53). However, the removal of copyrighted data from a large-scale training dataset may be impractical (54), given the large amounts of data and lack of documentation as described above (44). Moreover, it may be viewed as "undesirable" (52), as the inclusion of copyrighted works can be seen to strengthen the resulting model by providing a broader range of sources to draw on. It is significant that, as reported by Dehouche and Dehouche (7), the development of text-to-image models was supported by services such as *ArtStation* which "encourage artists to include detailed labels describing their work in order to make it more accessible to persons with disabilities," thus providing ready text-image pairs, but raising complex issues in relation to intellectual property when these images are used to train image generation models.

The tendency of image generation models to memorize and regenerate images ("image regurgitation" (55)) can pose a particular problem when the models are developed from "stolen artwork from the internet [*sic*] without permission". This has led to legal action, with a notable example being a lawsuit against Stable Diffusion by Getty Images, with output from Stable

Diffusion containing a regenerated version of the Getty Images watermark, overlaid on "grotesque" (56) images. There is an alternative perspective, however, which sees the regurgitation of training images as a positive, as it allows for the generation of specific images which have cultural significance (57).

An additional issue is how intellectual property rights should be handled for the works generated by AI, or the prompts used. Though the United States Copyright Office (58) only recognizes "'the fruits of intellectual labor' that 'are founded in the creative powers of the mind,'" it is as yet unclear how much human involvement would be required to meet this standard, or whether a well-articulated prompt, for example, might be sufficient (59,60). These issues have come to the fore given a growing market for prompt trading, where prompts are sold for generating art in particular styles (61,62). Though some conclude that the balance of arguments lies against granting copyright to machine-generated works (63), others argue that "[n]o AI acts alone, completely divorced from the influence of humans" (64). Thus, the rise of generative models brings into focus an urgent need to reconsider how copyright and other intellectual property rights are conceptualized (60).

### 3.3 Bias

A further issue in image generation models is bias (39,54,65). As described above, the proprietary and closed nature of image generation models poses a barrier to the evaluation of the full extent of algorithmic biases present in generative models (65). Though further work is urgently needed (65), preliminary work has begun to document how the output from image generation models reproduces and amplifies harmful stereotypes and biases (65–67). "Idiosyncratic biases along certain dimensions" (68) are reported in each of three major image generation models (DALL-E 2, Midjourney, and Stable Diffusion), though there are major challenges in ensuring that research on biases keep pace with the development of new or updated models.

The presence of intersectional biases in the models has been found to be a result of a lack of representativeness in the training data gathered from the Internet (69). They are also due to a Western and American-oriented bias in the models (70); there is thus a need to examine how bias may be reinforced through the use of other languages (71). Cho et al. (72,73) found that the use of image-text pairs from the Internet embedded intersectional biases between gender and skin tone in image generation models, finding this tendency reflected in output from models such as Stable Diffusion v1.4, which were more likely to generate images of certain gender and skin tone combinations even when prompts were neutral. Moreover, Cheong et al. (66) found that DALL-E Mini was "liable to reproduce, exacerbate, and reinforce extant human social biases," with a tendency to produce images that reinforced racial and gender stereotypes. Luccioni et al. (65) evaluated major image generation models, including Stable Diffusion v.1.4 and DALL-E 2, finding that both perpetuated stereotypical associations between race or gender and occupation, and did not accurately reflect their real-world distribution of genders and races in particular occupations—a tendency which has also been reported for DALL-E Mini (66).

Bianchi et al. (67) show that bias—including heteronormative biases—appear in images of people not just through the gendered and racialized depictions of individual figures in these images, but also through the way in which background objects such as houses or cars are rendered, finding that these contextual factors reproduce problematic stereotypes about intersections between multiple factors of identity including socio-economic status, race, and gender. There is also the risk of further amplification of bias through prompt modifications, as Struppek et al. (70) found that the use of even a single homoglyph in a prompt led to the generation of harmful stereotypes about non-Western cultures (48).

Bias in image generation models is of particular concern given the potential downstream applications of their output, as well as their possible impact on culture, as will be described below. These can include, as noted by Ungless et al. (48), representational harms through problematic representations of groups or individuals, as well as allocational harms through the distribution of opportunities or other goods. Output from image generation models is reported to be used not only in, for example, the generation of stock images which may then be widely used in a range of settings, but also in highly consequential settings, such as the creation of sketches of suspects in crime based on eyewitness accounts (65). There is also use in the medical domain, as Adams et al. (74) propose the use of DALL-E in generating images for education and training in radiology, even as their experiments with DALL-E 2 identified gender bias in the model.

As Fraser et al. (75) caution, the "worldview" of a particular model is not made visible to the user. Thus, "even when outputs appear biased, users' reactions are often to assume that the model made a one-off mistake and that its worldview truly aligns with their own." The authors suggest that "expert users" were aware of gender and racial biases in the models, but "novice users" often perceived the models as black boxed and thus were unaware of possible biased output, particularly as there are concerns that "the public's understanding of these models, from their development to their training data and their operational logics, is often limited" (76). This suggests that users may also unwittingly amplify bias through their personal use or promotion of problematic images. Yet, as Hutchinson et al. (77) note, "[i]t is impossible to avoid ambiguous inputs," given the nature of the visual modality. This creates additional space for bias to be introduced into the generated images, particularly as "underspecification" (77) remains a key issue. Although, as Offert and Phan (78) find, there have been attempts on the part of providers of text-to-image models to covertly modify prompts inputted by users to achieve more diverse outputs, current models are overly dependent on users to input prompts calling for diversity, whereas, they would ideally utilize an "Ambiguity In, Diversity Out" approach (75,77).

Finally, as Bianchi et al. (67) note, issues of bias are not easily mitigated, given that any understanding of an image is inherently subjective (79). Seeking to control the "worldviews," including biases, reflected in output from a particular model "requires considerable engineering time and effort"(75). Indeed, recent research has highlighted the way in which annotation can introduce bias into models (80), including through biases embedded in language itself (81), while even attempts to mitigate bias can in fact further reinforce it (82). Offert and Phan (78) draw attention to the fundamental "whiteness" of the output produced by models such as DALL-E 2, arguing that even more than bias, it is whiteness that is the "problem of foundation models" (see also (44,83–85). More broadly, Benjamin (83) and Joyce et al. (13) critique a tendency to seek out technical fixes to patch over issues of bias as a distraction from the urgent need to critically examine the systemic inequalities that produce bias on a societal scale. Thus, critical attention is needed to image generation models as potential site for the reproduction and amplification of biases.

## 3.4 Privacy

Given the opacity of training datasets, as delineated above, image generation models also raise privacy issues when training datasets include "privacy-sensitive information" (45) including facial images scraped from online sources. As discussed above, companies are disincentivized from making documentation on datasets—or the datasets themselves—publicly available for scrutiny, posing obstacles to addressing privacy concerns (45).

Diffusion models' image regurgitation tendencies run contrary to expectations that diffusion models would protect privacy (86). For example, though the use of diffusion models to generate synthetic training data has been proposed to mitigate privacy concerns, the tendency of diffusion models to regenerate the original images would obviate the benefits of such an approach (86). Moreover, privacy risks are heightened if captions for particular data samples in the training dataset are known, as users may prompt the generation of privacy-sensitive images, with membership leakage reported to be a "severe threat" in image generation (45,86). Struppek et al. (87) show that backdoor attacks on image generation models could potentially be used for the malicious creation of images of individuals, raising an urgent need for technical safeguards to protect against such risks (88).

These risks are heightened when the generated images are inherently sensitive, including facial images, which, as (89) has argued, are uniquely "central to our personal identities and social lives." Generative models may be used to generate problematic or compromising images of individuals by utilizing this capacity to generate facial images. Similarly, Van Le (88) draws attention to Textual Inversion and DreamBooth as two techniques commonly used to generate personalized output, but which are liable to be abused to generate harmful images. There have been reports of the presence of medical images of individuals in the LAION dataset, as well as of radiological images including CT, MRI, and ultrasound images in data for DALL-E 2 (74). Ultimately, Zhang et al. (54) suggest that it is "nearly unfeasible to fully address" privacy concerns, though it is noteworthy that some efforts have been made to offset these issues such as through OpenAI's efforts to deduplicate training data for DALL-E 2 to reduce images replicated in output (55,90).

## 3.5 Impact on the informational environment

As has been reported for LLMs (91), the output of generative models can have a profound impact on the informational environment over the longer term, posing "severe threats to society" (92,93). Here, we define the informational environment broadly, including cases in which generated images are harmful or damaging. Thus, there is significant overlap between the issues identified here and those coming under the umbrella of prviacy in the section above.

There are concerns particularly about the potential for highly realistic images and deepfakes generated through image generation models to harm the informational environment (92,93). This is in part as image generation establishes new norms through which "audio and video content cannot be taken at face value," thus threatening trust in the media, and exacerbating shifts towards post-truth societies (94). There are early signs that these shifts may already be underway, as Oppenlaender et al. (95), in a small-scale, exploratory online survey of public perspectives on the use of image generation models, found that participants were already concerned about the "use of AI-generated imagery for opinion manipulation, fake news, and `deep fakes [*sic*].'"

Furthermore, though their study focused on images synthesized by GANs, Nightingale and Farid (96) found that "synthetically generated faces are not just highly photorealistic, they are nearly indistinguishable from real faces and are judged more trustworthy," emphasizing that such artificial images carry "serious implications for individuals, societies, and democracies." They highlighted the risk that such images could be used for "nonconsensual intimate imagery, fraud, [and] disinformation campaigns"(96). This includes harmful child sexual abuse material (CSAM), as noted by Bird et al. (10) and by the Internet Watch Foundation (97) which recently found "20,254 AI-generated images…posted to one dark web CSAM forum in a one-month period." Though these risks are not entirely novel (98), they can be expected to be amplified through the ease of access to, the "democratization" (40) of, and the increasing public awareness of image generation systems, as "high-quality fakes now seem to come out from an assembly line." Similarly, Agarwal et al. (99) call for urgent attention to the risks of AI-synthesized imagery, which they argue can "pose a significant threat to our democracy, national security, and society," which is compounded by the likelihood that such synthesized images could be rapidly spread (94). This occurs in part through the "liar's dividend," through which "public figures caught in genuine recordings of misbehavior will find it easier to cast doubt on the evidence against them" (94). Ricker et al. (93) "find that existing detectors, which claim to be universal, fail to effectively recognize images synthesized by different [diffusion models]," highlighting an urgent need for mitigation of these risks, including through mechanisms for improved detection (see also (98)), or mechanisms to make the provenance of AI-generated output clear, such as through the use of watermarks (59,96).

The generation of more explicitly harmful images also poses a particular risk to the informational environment. Qu et al. (100) identify five kinds of unsafe images, including sexually explicit, violent, disturbing, hateful, and political images, finding through their study that just under 15 percent of all analyzed images generated through a text-to-image model could be classed into these categories, despite attempts to restrict the output of harmful content. In addition, Struppek et al. (87) showed that backdoor attacks on image generation models could be injected within two minutes and used to generate harmful images, with even "a single violent or explicit image" potentially causing harm to the viewer. This highlights the potential vulnerabilities that can be exploited to cause further harm to the informational environment, beginning at the level of the individual.

## 3.6 Impact on the cultural environment

Image generation models further pose risks to culture. Srinivasan and Uchino (101) identify three key cultural risks through the use of image generation models: first, as described above, the biases embedded in image generation models risk causing harm to the cultural environment; second, the models allow users to stereotype the styles of artists or movements, while failing to convey the intent of the original artists, thus threatening cultural heritage; and third, through the generation of inaccurate or unrepresentative images of historical events, they impede historical and cultural preservation. These issues are compounded by imbalances in the art styles drawn on in training data, such as an over-representation of Renaissance and other European art, or of modern art (101), with biases amplified by the Western contexts in which the models are developed (57).

The use of image generation models in art generation may limit creativity, by generating "predictable" (40) images, and impeding personalization. Yet, the longer-term impact of generative models on the cultural environment is often discussed with reference to photography; it transformed how art was produced and appreciated, suggesting that image generation models may, in a similar way, ultimately become a tool to enhance the work of artists (53). These uses may include identifying reference images, speeding up prototyping processes, reducing bias in art, and empowering artists to work with new software or with new media (7,40). Nonetheless, at present many artists are "perplexed and disoriented" (40) regarding how to implement AI tools in their art, and there is also a growing concern over the implications of the replacement of human artists by image generation models (102), as employment opportunities, particularly for freelance artists, may be lost if such models are utilized instead of humans (53). This is problematic given that artists' works are used to train models, as described above, yet the profit from any such training remains with the companies providing the models, ultimately creating a situation in which companies are "profiting off the artistic endeavors of others, and possibly supplanting those human artisans" (53).

These impacts may extend to education in the area of the visual arts (7) and the potential impact on the next generation of artists is particularly of concern (103). Oppenlaender et al. (95) found that participants in their study were concerned about the impact of generative models on culture, including concerns over a "decline in human creativity," suggesting that this is an issue which members of the public are already attuned to. However, Oppenlaender (104) has argued elsewhere that current cultural understandings of creativity are overly limited by a focus on product, and thus overlook the role of creativity which arises from the process of image generation using text-to-image models, which includes human involvement, thus complicating understandings of the implications of the generative models for the arts

In this vein it is also significant that Ragot et al. (105) found confirmation of a tendency to evaluate human-created artwork more highly than works created by generative models. They suggest that negative bias towards artwork generated by models may be due to intergroup bias, through which "AI may be anthropomorphized and thus considered as the outgroup," drawing on psychological research which suggests that individuals may have negative biases towards outgroups. Nonetheless, AI-generated art has been awarded prestigious prizes, including cases where the judges were unaware of the use of image generation models (103,106).

### 3.7 Impact on the natural environment
In the context of growing concern over the breach of planetary boundaries (107–109), there is an urgent need to consider the extractive processes behind generative models and their environmental impact (108,110,111). Van Wynsberghe (112) has drawn attention to concern over the sustainability of AI as a third wave of ethics, succeeding second wave concerns about bias and fairness. Brevini (111) has called for urgent action on "quantifying and reducing the environmental costs and damages of the current acceleration" of AI development. There has to date been relatively limited attention to the environmental costs of generative models. However, concerning LLMs, for example, Bender et al. (41) notably highlight environmental costs as one of the primary concerns in pursuing their development. Despite this, Rillig et al. (113) report that, even two years after the initial call from Bender et al. "the debate so far largely missed the potential implications of current and future LLM tools for the environment," a trend which we find to be similarly true for text-to-image models. Yet, Luccioni et al. (114) indicate that "[t]asks involving images are more energy- and carbon-intensive compared to discriminative tasks," with image generation as one of the most intensive of these tasks—highlighting the urgency of these considerations.

In 2018, OpenAI (115) announced that "the amount of compute used in the largest AI training runs has been increasing exponentially with a 3.4-month doubling time," leading to an increase of 300,000 times since 2012—a figure which would have been 7 times if aligned with Moore's Law's two-year doubling period. This highlights the growing environmental burden arising from the training of large generative models, and particularly through increased carbon emissions. Utilizing the Machine Learning Impact Calculator (116) and based on an analysis of hardware, use times, cloud providers, and region, the estimated emissions from the development of Stable Diffusion v1 were equivalent to 15,000 kg of carbon dioxide (3).

It is noteworthy, however, that emissions are just one aspect of the environmental burden from image generation models, and little data has been made available about other aspects (108,114,117). These aspects are varied and include the use of significant amounts of freshwater to cool data centers, as reported for LLMs, even at a time when shortages of freshwater are growing increasingly worrisome, and major tech companies have committed to reducing their "water footprints" (118). They also include the costs inherent in the needed hardware and infrastructure across their life cycles, ranging from the extractive processes involved in metals and minerals used in the systems such as rare earth elements, through to their end-of-life as e-waste (108,119,120). These processes cause major harm directly and indirectly to both human and environmental health (119). As increasingly advanced generative models are designed and implemented, there are suggestions (121) that new devices to operate these models will also need to be concurrently made available – further heightening these environmental costs, and leading to increased e-waste, with its attendant impacts on health (108,119).

# 4. Discussion

Thus, we have brought together the technical and social perspectives on the life cycle of image-generation models. Here, we situate the social issues arising from image generation models within their life cycle. We also compare these issues in image generation models with those of LLMs to argue that there is an urgent need for consideration of the risks of image generation models at minimum commensurate with the focus that has been given to LLMs.

## 4.1 Social issues in the model life cycle

This section examines the intersection of the technical and social axes, situating the social issues addressed above within the model life cycle, thus we review each issue cluster identified in the results above, and assess their relevance within the life cycle. As will be clear through the discussion below, each issue cluster is relevant across the model life cycle, though its relevance may be heightened in particular phases.

Data issues appear across multiple sub-phases of the model life cycle. They are intrinsic in current data acquisition procedures for text-to-image models. They also appear through model design, particularly when the models are trained and later refined. These issues include the involvement of human in model refinement under exploitative conditions. These issues are then inevitably reflected in the output from the models. For example, the lack of documentation of the content of datasets and the challenges in pre-emptively identifying and tackling risks means that the generated output contains problematic content. As discussed above, if these generated images are scraped and used in further training, these issues recur cyclically in the next generation of models.

Intellectual property issues likewise appear similarly across multiple sub-phases of the model life cycle. Issues arise both through the data which is used for the models, which may include works in violation of intellectual property rights, but can also appear in output, as when works generated by a text-to-image model are overly similar to existing works. Although as yet unreported, there may also be platform-based issues if platform owners skim and later utilize inputted prompts. These issues also extend into the deployment phase, as debates have arisen regarding the handling of intellectual property in prompt generation and input.

In relation to bias, the training data used often reflects structural and embedded social biases present in the sources from which the data is scraped. Then, these biases are embedded in training, and can also be reinforced through the refinement process particularly if humans are involved in this process and their own personal biases are reflected. Bias can also become pertinent in the input process, as Offert and Phan (122) identified a tendency in DALL-E 2 for prompts to be modified unbeknownst to the user in order to generate more diverse images. Once these images are outputted, bias risks being further reproduced through the use of these problematic images.

Privacy-sensitive images may be used in data acquisition and training, heightening the relevance of privacy concerns in this phase. Additional issues arise in the deployment phase if users utilize the text-to-image models to attempt to generate privacy-sensitive images, such as explicit images or images depicting specific individuals. Moreover, as with intellectual property issues above, there is a risk that platform owners skim prompts inputted by users. The spread of these generated images, and their potential re-use in further models pose additional privacy risks.

Issues in the informational environment can arise through data acquisition as there is no way to assess the accuracy or factual nature of the image-text pairs acquired. These issues then come to the fore in deployment, where the models can be used to intentionally or inadvertently generate images which misrepresent reality. Without watermarking or other indications through which to clarify the provenance of these images, there are key concerns about their detrimental impact on the informational environment at large once they are used and are in circulation.

As with intellectual property issues, cultural issues occur when culturally important images are scraped for use by the model. The issues similarly re-emerge in deployment and use of the images, such as when the generated images misrepresent culturally important events or works. They also pose broader cultural issues when the platforms are utilized to replace the work of human artists.

And finally, issues related to the natural environment are unique in this analysis as they emerge across all phases of the model life cycle. Each step in the model life cycle implicitly involves an environmental cost, as each piece of hardware and all infrastructure that is used in the development and deployment of text-to-image models create burdens and pressures on the natural environment. Though there remains a dearth of relevant research in this area, these are nonetheless critical considerations.

As is clear through the discussion above, the social issues of image generation models appear across the model life cycle. For this reason, it is essential that pre-emptive consideration of these issues be embedded throughout all phases of the model life cycle. In Table 2 below, we propose a matrix which could be used in risk assessments of present and future image generation models at the intersection of the social and technical axes.

*Table 2* Proposed risk assessment matrix

| Phase | Data | IP | Bias | Privacy | Information | Cultural | Natural |
|---|---|---|---|---|---|---|---|
| **Data acquisition** | | | | | | | |
| Image | | | | | | | |
| Text | | | | | | | |
| Filtering | | | | | | | |
| **Model design** | | | | | | | |
| Model type | | | | | | | |
| Training | | | | | | | |
| Refinement | | | | | | | |
| **Deployment** | | | | | | | |
| Input | | | | | | | |
| Platform | | | | | | | |

## 4.2 Comparison with LLMs

Recent research such as by Bender et al. (41) and Weidinger et al. (12) have highlighted the risks associated with LLMs. Yet, despite their growing utilization and societal influence, relatively little comparable evaluation of image generation models has been conducted, though there are first movements in the field (10,11). Here, we compare the issues arising from LLMs, as reported in prior research, with the issues we have identified through this study, considering each of the seven issue clusters along the social axis, finding levels of risk across multiple issue clusters that meet or exceed those of LLMs described above.

Beginning with data issues, just as with LLMs, image generation models rely on massive, undocumented datasets, which means that both models and their datasets are black-boxed. This poses obstacles to the identification of possible sources of harm and to their remediation. Furthermore, though issues around attempts at content moderation have yet to receive as much attention as for LLMs (51), the documented trauma of content moderators (50) for other types of image-based content suggest that these issues can be expected to be similarly present for image generation models.

Intellectual property issues arise with both LLMs and image generation models, yet the spate of recent lawsuits around image generation models—which have yet to be matched by a comparable level of legal activity around LLMs—suggest that issues are heightened in this area. Indeed, though LLMs may potentially produce identical text to a copyrighted work, similarities on the level of style rather than exact replication may be harder to detect and more contestable when the medium is text; however, the similarities between images—particularly given diffusion models' tendency to regenerate training data—are more readily identifiable, foregrounding intellectual property issues.

Bias is an issue which arises with both LLMs and image generation models. Yet, as discussed in above, particularly with reference to work by Bianchi et al. (67) image generation models include contextual information beyond what is in a more straightforward medium such as text; thus, bias can be reproduced not only in the foregrounded object in an image, but also in the background context, and it can be difficult for users to provide sufficient specficity in prompts to avoid this reproduction of bias. Here, then, the potential for bias to manifest in output is amplified in comparison with LLMs.

Privacy issues arise similarly in LLMs and image generation models, as both may draw on personal information about individuals which has been posted online. In relation to LLMs, there are risks, for example, that individuals input sensitive information into LLMs, and that this data be leaked or otherwise reflected in output towards other users, as was reported for ChatGPT (123). However, privacy risks are heightened in image generation models given that they risk the regurgitation of training data, including facial images, which are inherently sensitive. Of particular concern in relation to privacy are reports of sensitive medical images present in the datasets for image generation models.

The impact on the informational environment is particularly severe in the case of image generation models. Though it is well-documented that LLMs may be used for disinformation campaigns and to flood the informational environment with inaccurate or misleading information (e.g.,(91)), images evoke visceral responses in their viewers (124), and are often even more likely to be taken for granted as true than are text (see (94)). In comparison with text, it is a relatively recent development that images can be faked, and thus the cultural sensitivity towards the need to question visual evidence is as yet underdeveloped, further heightening the risks in this area. Indeed, images are "remarkably slippery things, laden with multiple potential meanings, irresolvable questions, and contradictions" (79).

Similarly, the impact on the cultural environment is also notable in image generation models, as reported by (101). The impact of image generation systems extends beyond just the future of art and imagery, to also encompass their impact on how the past is understood, if image generation models are used to generate images of historical events. Moreover, they risk bringing about fundamental changes to how art and creativity are produced and understood, and to the livelihoods of artists.

The impact of LLMs and image generation models on the natural environment hold similarities in that both are insufficiently understood, as there remains insufficient documentation of their impact across their life cycles, including both software and hardware. Given the relatively larger amount of data which is drawn on for LLMs in comparison with image generation models, there are indications that the impact on the natural environment may be more pronounced in the case of LLMs. Nonetheless, further research is urgently needed to better understand these impacts.

Thus, across the majority of these issue clusters, image generation models pose risks comparable to those of LLMs.

## 4.3 Recommendations

Although mitigations are outside the scope of this exploratory analysis, a number of recommendations arose from this interdisciplinary work and are included here. These are grouped into three categories: mitigations by design, mitigations post-hoc, and general recommendations. Mitigations by design are those which should be considered during the development phase of the models and involve pre-emptive consideration of the issues arising from each issue cluster. Mitigations post-hoc aim to mitigate already existing issues in the models. Finally, the general recommendations are applicable across the phases of the model life cycle.

1) **Mitigations by design**:
   a. To be aware of data imbalances and to increase diversity in datasets used for training generative models, to offset issues related to bias and the impact on the cultural environment. It is noteworthy that, although as discussed above, imbalances in text-to-image models' datasets have been reported (125,126), there are insufficient resources to effectively address them at present.
   b. To assess model performance not only in terms of accuracy or image quality, but also in terms of factors directly related to the issue clusters, such as bias or environmental impact.
   c. To create standardized benchmarks to this end.
2) **Mitigations post-hoc**:
   a. To put increased effort into technical mitigation strategies such as bias mitigation or "machine unlearning" (127–129). However, techniques that target a single issue may not be optimal. For example, whereas machine unlearning can help with copyright and privacy issues, it may worsen the representation of minorities (130).
   b. To make datasets and models publicly available to allow for external audits by individuals and institutions, to facilitate identifying and addressing emergent issues.
3) **General recommendations**:
   a. Mitigations should be undertaken with awareness of their possible impact on other issue clusters. As we have shown here, each phase of the model life cycle can involve multiple issues, with overlapping effects.
   b. Given the limitations to a "soft" (131), researcher-driven approach to ethical technology development, we call for further regulation on image generation models that is grounded in currently observed and anticipated risks along the seven issue clusters identified. Although there are attempts within the global community to begin to regulate AI such as through the European Union's proposed AI Act and the G7 Hiroshima AI Process, there has been notable unevenness in the consideration of these issue clusters, with significant attention to data and intellectual property clusters, with less concern over, for example, the impact on the natural environment. It is essential that future, comprehensive regulation address all seven of the issue clusters.

**4.4 Final remarks**
We have synthesized the literature on the social issues posed by image generation models, identifying seven key issue clusters: data issues, intellectual property, bias, privacy, and impacts on the informational, cultural, and natural environments. We have then intersected these considerations along the social axis with those along the technical axis, highlighting where these risks may arise along the model life cycle. Furthermore, through comparison with the literature on LLMs, this study has shown that the risks associated with image generation models at least match, if not exceed, the risks posed by LLMs, though image generation systems have received significantly less critical attention to date.

Given the rapid pace of development of generative models, the limited time frame and scope is an inherent limitation of this study. This work should be updated and expanded to capture further developments in the field. Moreover, if the development and deployment of image generation models are to be pursued at the current frenetic pace, this work indicates that there is an urgent need to consider how these issues and risks can be mitigated. An in-depth exploration of existing mitigations for each of the identified issue clusters was beyond the scope of this study, but future work must examine this. To this end, we highlight here the merits of sociotechnical considerations grounded in interdisciplinary collaboration, drawing together the on-the-ground perspectives of technical experts with the insights of experts in the social sciences.

## 6. Acknowledgments

The authors wish to thank Dr. Yusuke Shikano, Keiichiro Suzuki, and Yusuke Nagato for their participation in the development of this study.